\documentclass[nofootinbib,twocolumn,showpacs,aps,10 pt,pra]{revtex4-1}
\usepackage{graphicx}
\usepackage{amsmath}
\usepackage{amssymb}
\usepackage{amsthm}
\usepackage{multido}
\usepackage{psfrag}
\usepackage{verbatim}
\usepackage{xcolor}

\newcommand{\tr}{\mathrm{tr}}

\newcommand{\M}{\mathcal M}

\renewcommand{\P}{\mathcal P}

\renewcommand{\th}{\mathrm{th}}
\newcommand{\D}{\mathcal D}

\newcommand{\cl}{\mathrm{cl}}
\newcommand{\letter}{work~} 

\begin{document}
\title{Phase space formalism for quantum estimation of Gaussian states}
\author{Alex Monras}
\affiliation{Centre for Quantum Technologies, National University of Singapore, 2 Science Drive 3, 117542, Singapore}
\begin{abstract}
We formulate, with full generality, the asymptotic estimation theory for Gaussian states in terms of their first and second moments. By expressing the quantum Fisher information (QFI) and the elusive symmetric logarithmic derivative (SLD) in terms of the state's moments (and their derivatives) we are able to obtain the noncommutative extension of the well known expression for the Fisher information of a Gaussian probability distribution. Focusing on models with fixed first moments and identical Williamson 'diagonal' states --which include pure state models--, we obtain their SLD and QFI, and elucidate what features of the Wigner function are fundamentally accessible, and at what rates. In addition, we find the optimal homodyne detection scheme for all such models, and show that for pure state models they attain the fundamental limit.
\end{abstract}


\maketitle

Estimation theory plays a central role in modern developments of quantum enhanced metrology. From a practical point of view, it allows to assess the ultimate precision limits of given metrological schemes. From a fundamental perspective, it provides a gold standard upon which to asses distinguishability of quantum states. Quantum estimation theory is an old subject~\cite{helstrom_quantum_1976,holevo_probabilistic_1982} and it has seen huge developments over the last 20 years~\cite{braunstein_statistical_1994,boixo_generalized_2007,giovannetti_advances_2011}. It has played a major role in understanding the fundamental powers and limitations of quantum measurement. The groundbreaking advances in atomic clocks and precision metrology~\cite{appel_mesoscopic_2009,ma_quantum_2011} suggest that quantum estimation theory will only become more relevant as measurement precision reaches its fundamental limits.

Common to almost all the disciplines of physics where precision metrology can provide significant results, is the fact that they benefit from the simplicity and power of the Gaussian state formalism~\cite{weedbrook_gaussian_2012}. The latter has already proven its success and serves as an invaluable tool in describing quantum states of light and atomic ensembles, as well as providing useful insight and intuition. 

Despite the great success of both the Gaussian state formalism, and quantum estimation theory, these two have never been successfully merged. Indeed, the nontrivial equations defining central objects in quantum estimation theory often forces to numerical methods~\cite{adesso_optimal_2009,dorner_optimal_2009,demkowicz-dobrzanski_quantum_2009} for computing precision bounds and determining optimal measurements, and these difficulties are only aggravated by the infinite-dimensional nature of bosonic systems. Most remarkably, these difficulties are not alleviated by the Gaussian state formalism, but for very particular cases and without explicit harvest of the phase-space structure. Ironically, it is in the seminal book by Holevo~\cite{holevo_probabilistic_1982} where one encounters one of the first comprehensive accounts of the phase-space formalism and Gaussian states on the one hand, and the foundations of quantum estimation theory on the other.  There, certain essential features of Gaussian states regarding optimal detection were established, but the analysis focused on the Gaussian shift model. This model, despite being extremely relevant, and certainly the first candidate to be studied, is not well suited for modern entanglement-enhanced metrology, where the signal is encoded in the state's correlations rather than the amplitudes. It is the purpose of the present \letter to provide a fully general phase-space formulation of the central quantities in quantum estimation theory, namely, the \emph{symmetric logarithmic derivative} (SLD), and the \emph{SLD quantum Fisher information}, with focus on general Gaussian states. After presenting this new formulation, we use it to study a wide class of models which include all pure state models. We also address the optimality of Gaussian measurements and derive sufficient conditions under which they are optimal.

Given a quantum model, \emph{i.e.} a parametrized set of quantum states $\mathcal S=\{\rho_\theta\}$, the quantum Cram\'er-Rao bound (QCRB) establishes a lower bound to the variance of any unbiased estimator of~parameter~$\theta$,
\begin{align}
	(\Delta\hat\theta)^2\geq I_Q(\theta)^{-1}
\end{align}
where $I_Q(\theta)$ is the SLD quantum Fisher information (QFI). The QFI is defined in terms of the symmetric logarithmic derivative (SLD), which is the Hermitian operator $\mathcal L_\theta$ that satisfies
\begin{align}\label{eq:SLD}
	\partial_\theta\rho_\theta=\frac{1}{2}(\rho_\theta \mathcal L_\theta+\mathcal L_\theta\rho_\theta),
\end{align}
and the QFI reads
\begin{align}\label{eq:Fisher}
	I_Q(\theta)=\tr[\rho_\theta \mathcal L_\theta^2].
\end{align}
Although the QCRB only establishes a bound, for uniparametric models it is asymptotically attainable~\cite{gill_state_2000,gill_asymptotics_2001,paris_quantum_2009}. Therefore, it establishes the best asymptotic rate at which statistical fluctuations can decrease when measuring parameter $\theta$~\cite{helstrom_quantum_1976,holevo_probabilistic_1982,braunstein_statistical_1994}. Thus, it is a quantitative measure of distinguishability of a state $\rho_\theta$ from its neighbors $\rho_{\theta\pm\delta\theta}$ and as such, is intimately related to the quantum Fidelity and the Bures distance~\cite{bengtsson_geometry_2006}. On the other hand, the SLD not only has a geometric meaning; it also represents, by construction, an optimal observable~\cite{gill_asymptotics_2001,paris_quantum_2009}, in the sense that --to leading order-- $\langle \mathcal L_\theta\rangle_{\theta+\delta\theta}$ is proportional to the deviation from a reference state $\rho_\theta$, $\delta\theta$, and it minimizes the statistical fluctuations.

We focus our analysis on systems of $n$ bosonic modes, described by Hilbert space  $\mathcal H={\rm L}^2(\mathbb R^n)$. These systems are characterized by $2n$ canonical variables ${R^i=(Q_1,\ldots,Q_n,P_1,\ldots,P_n)}$ with canonical commutation relations $[R^i,R^j]=i\omega^{ij}$, where $\omega$ is the $2n\times2n$ symplectic matrix, and $\Omega$ is the degenerate symplectic inner product, $\Omega(a,b)=\sum_{ij}a^ib^j\omega^{ji}=-\Omega(b,a)$\footnote{Summation of latin indices $i$ and $j$ runs from 1 to $2n$, except for symbol $k$ which labels the $n$ modes and thus runs from $1$ to $n$}. The Weyl (displacement) operators are defined as ${W(\xi)=\exp i\Omega(\xi,R)}$, so that $W^\dagger(\xi)R^iW(\xi)=R^i+\xi^i$, and $W(\xi)W(\eta)={W(\xi+\eta)e^{\frac{i}{2}\Omega(\xi,\eta)}}$. In addition, we introduce the symmetric product $A\circ B=(AB+BA)/2$. The symmetric product is not associative. We define it to have precedence over ordinary product, $A\circ B\,C=(A\circ B)C\neq A\circ (BC)$.

Gaussian states are defined as those states $\rho\in\mathcal B(\mathcal H)$ having Gaussian characteristic function $\chi_\rho(\xi)=\tr[\rho W(\xi)].$ Let $\M_{2n}$ be the space of $2n\times2n$ matrices over $\mathbb R$. We define the first and second moments as
\begin{subequations}
\begin{align}
\label{eq:firstmoment}
	d^i&=\tr[R^i \rho]\\
\label{eq:secmoment}
	\Gamma^{ij}&=2\tr[(R^i-d^i)\circ (R^j-d^j)\,\rho],
\end{align}
\end{subequations}
where $d\in\mathbb R^{2n}$ and $\omega, \Gamma\in \M_{2n}(\mathbb R)$. With these definitions, a Gaussian characteristic function reads
\begin{align}\label{eq:chi}
	\chi_\rho(\xi)=\exp\big(i\xi^\top \bar d-\tfrac14\xi^\top \bar\Gamma\xi\big)
\end{align}
where $\bar d =-\omega d$ and $\bar \Gamma=\omega\Gamma\omega^\top $.

The main goal of this \letter is to provide a formulation of Eqs.~\eqref{eq:SLD} and \eqref{eq:Fisher} in terms of $d_\theta,\Gamma_\theta,\partial_\theta d_\theta$ and $\partial_\theta \Gamma_\theta$, and to explore the benefits of such formulation. Let $\mathcal S=\{\rho_\theta\}$ be a Gaussian model with parameter $\theta$. Any such model is fully described by the first and second moments $d_\theta$ and $\Gamma_\theta$ of $\rho_\theta$. As can be expected from the structure of Gaussian states, --and we show in Appendix \ref{app:SLD}-- the SLD is quadratic in the canonical operators. It has also zero expectation as follows from its definition Eq.~\eqref{eq:SLD} and $\partial\tr[\rho_\theta]=0$. Taking as ansatz for $\mathcal L_\theta$ the expression
\begin{align}
\nonumber
	\mathcal L_\theta=&\sum_{ij}L_{ij} (R^i-d^i)\circ(R^j-d^j)\\
\label{eq:SLDfinal}	
&+\sum_i b_i (R^i-d^i)-\frac{1}{2}\tr[L\Gamma]
\end{align}
in Eq.~\eqref{eq:SLD} and taking the characteristic function of both sides one  can relate the derivative of $\chi_\rho(\xi)$ w.r.t to $\theta$, $\partial_\theta \chi_\rho(\xi)$ (lhs) to expectations of $R^i\circ W(\xi)$, and $(R^i\circ R^j)\circ W(\xi)$ (rhs). This relation in turn, implies a relation between $d,\Gamma,\partial d$ and $\partial\Gamma$ (lhs) and suitable linear combinations of $\partial_{ij} W(\xi)$ (rhs), where $\partial_{i}\equiv\partial/\partial\xi^i$. Equating terms with equal powers of $\xi$ yields solutions to $L$ and $b$ in Eq.~\eqref{eq:SLDfinal}. We report the solutions here and the technical details 
in Appendix~\ref{app:SLD}. Define the linear map $\D_X:\M_{2n}\rightarrow \M_{2n}$ as
\begin{align}
	\D_X(Y)&=X Y X^\top-\omega Y \omega^\top.
\end{align}
With this, $L\in\M_{2n}$ and $b\in\mathbb{R}^{2n}$ are given by
\begin{subequations}\label{SLDparams}
\begin{align}
	b&=2\Gamma^{-1}\partial d,\\
\label{eq:Dinverse}
	L&=\D^{-1}_\Gamma(\partial\Gamma),
\end{align}
\end{subequations}
where the a pseudoinverse (Moore-Penrose inverse) is understood whenever $\D_\Gamma$ is singular. The map $\D_\Gamma^{-1}$ in Eq.~\eqref{eq:Dinverse} will play a central role in our discussion. Notice that $\D_\Gamma(X^\top)=\D_\Gamma(X)^\top$ is symmetry preserving. Let $Y=\D_\Gamma^{-1}(\partial\Gamma)$. Then $Y$ satisfies the Stein equation, $Y-F Y F^\top=-\partial \Gamma^{-1}$, where $F=(\Gamma\omega)^{-1}$ and its spectrum is contained in the unit disk ($|F|\leq1$). Assuming $\rho$ is nonsingular, $|F|<1$ and the unique solution to the Stein equation is~\cite{bhatia_positive_2006}
\begin{align}\label{eq:solution}
	Y=-\sum_{k=0}^\infty F^k \partial \Gamma^{-1} F^\top{}^k.
\end{align}
A more detailed analysis of the structure of $\D_\Gamma$ and its pseudoinverse is given in Appendix \ref{app:SLD}. The following relation will be useful,
\begin{align}\label{eq:DDinv}
	\partial\Gamma=-\D_\Gamma(\partial\Gamma^{-1})-\omega\partial\Gamma^{-1}\omega^\top,
\end{align}
as follows from $\partial\Gamma=-\Gamma\partial\Gamma^{-1}\Gamma$.

As a first observation regarding the structure of the SLD obtained in Eq.~\eqref{eq:SLDfinal}, setting $\partial\Gamma=0$ in Eq.~\eqref{eq:Dinverse} recovers the Gaussian shift model, already studied extensively in the literature \cite{holevo_probabilistic_1982,gill_asymptotic_2012}. In this case we obtain $\mathcal L_\theta=\partial d\cdot\Gamma^{-1}\cdot R$, linear in the canonical operators, hence recovering the optimality of homodyne detection.

Beyond the Gaussian shift model, for generic Gaussian models the SLD is at most quadratic in the canonical operators. Defining operators $\hat R^i=R^i-d^i+\tfrac12[L^{-1}b]^i$ one can write
\begin{align}
	\mathcal L_\theta=\sum_{ij}L_{ij} \hat R^i\circ \hat R^j+C
\end{align}
where $C$ is a scalar. In addition, as follows from Eqs.~\eqref{eq:Dinverse} and \eqref{eq:solution} the matrix $L$ is the image of $-\partial\Gamma^{-1}$ under the action of a completely-positive map. Therefore, whenever $\partial\Gamma^{-1}$ or $-\partial
\Gamma^{-1}$ is positive semidefinite, $L$ is so too, and there is a symplectic transformation $T_\theta$ such that $L=T_\theta^\top \mathrm{diag}(\alpha_1,\ldots,\alpha_n,\alpha_1,\ldots,\alpha_n) T_\theta$. Thus, defining $\hat R^i=\sum_j T^{ij} R^j$ and the number operators, $N_k=\tfrac12[(\hat R^k)^2+(\hat R^{n+k})^2-1]=\tfrac12[(\hat Q^k)^2+(\hat P^k)^2-1]$ one has
\begin{align}
	\mathcal L_\theta= \sum_{k=1}^n \alpha_k (N_k-\langle N_k\rangle_\theta).
\end{align}
Therefore, in the case where $\partial\Gamma^{-1}$ has definite signature, the SLD reduces to photon counting in suitably defined modes. This result extends and generalizes the findings of \cite{monras_optimal_2007, monras_information_2010}, where the SLD was shown to have this structure for certain classes of channel estimation problems. Still, despite having a physical interpretation of the operator $\mathcal L_\theta$ (Gaussian transformations and photon counting), implementing a measurement of it may still be prohibitive, especially if strong squeezing of the signal is required.

Before considering more practical measurements, let us obtain the fundamental limit to their performance. From Eq.~\eqref{eq:SLDfinal} one can readily obtain the expression for the quantum Fisher information for general Gaussian models. It is convenient at this point to endow $\M_{2n}$ with an inner product structure, $(X|Y)=\tr[X^\top Y]$, thus regarding matrices as vectors (kets) $|X)$, and linear maps thereof as $A\otimes B|X)\equiv |A X B^\top)$, so that we can write $\D_\Gamma=\Gamma\otimes\Gamma-\omega\otimes\omega$ and Eq.~\eqref{eq:Dinverse} reads $|L)=\D_\Gamma^{-1}|\partial\Gamma)$. With this inner product, $\mathcal D_\Gamma$ and its (pseudo)inverse are self-adjoint. The QFI reads [see Appendix \ref{app:QFI} for details]
\begin{align}\label{eq:QFI}
	I_Q
			=\frac{1}{2}(\partial\Gamma|(\Gamma\!\otimes\!\Gamma-\omega\!\otimes\!\omega)^{-1}|\partial\Gamma)+2\partial d^\top\Gamma^{-1} \partial d.
\end{align}
This expression is, for the first time, the most general form of the QFI in the Gaussian state formalism, and together with Eqs.~\eqref{eq:SLDfinal} and \eqref{SLDparams} constitutes our main result. It allows to compute precision bounds for a number of situations, and expresses the Fisher information in a form amenable for numerical computation, overcoming the difficulties posed by the infinite-dimensional character of Eq.~\eqref{eq:SLD}. Notice that putting the dimensions back in and taking the classical limit ($\hbar\rightarrow0$) yields $\D_\Gamma^{-1}=(\Gamma\otimes\Gamma-\hbar^2\omega\otimes\omega)^{-1}\rightarrow \Gamma^{-1}\otimes \Gamma^{-1}$, thus recovering 
\begin{align}\label{classicalI}
	I_{\cl}=\frac12\tr[\partial\Gamma \Gamma^{-1}\partial\Gamma\Gamma^{-1}]+2\partial d^\top\Gamma^{-1}\partial d, 
\end{align}
the Fisher information of a Gaussian probability distribution centered at $d$ with covariance $\Gamma$ \footnote{The factor of 2 in the second summand is due to the factor 2 in the definition of the covariance matrix, Eq.~\eqref{eq:secmoment}}. Thus, Eq.~\eqref{eq:QFI} is the noncommutative generalization of Eq.~\eqref{classicalI}. Indeed, $\D_\Gamma^{-1}$ plays an essential role in capturing the geometry and distinguishability properties of Gaussian states, and the term proportional to $\hbar^2\omega\otimes\omega$ accounts for the uncertainty due to noncommutativity of the canonical observables.

Eq.~\eqref{eq:solution} gives the unique solution for models with nonsingular $\rho$, namely, no vacuum modes in the Williamson decomposition. This rules out the important case of pure state models. However, a generic solution to the Stein equation can be given for the class of models with covariance matrix satisfying the relation 
\begin{align}\label{eq:Gibbs}
	(\Gamma\omega)^2=-\nu^2,
\end{align}
where $\nu\propto I$ is constant and we will treat it as a scalar. We call these models \emph{isothermal} due to the constant temperature of their Williamson decomposition. These models include all pure state Gaussian models, as well as some mixed models that have recently attracted attention~\cite{blandino_homodyne_2012}. It is easy to check that Eq.~\eqref{eq:Gibbs} implies $(\nu^2-1)\Gamma=\nu^2\D_\Gamma(\Gamma^{-1})$, which combined with Eq.~\eqref{eq:DDinv} leads to $(1+\nu^2)\D_\Gamma^{-1}(\partial\Gamma)=-\nu^2\partial\Gamma^{-1}$. Despite some technicalities in simplifying $\D^{-1}_\Gamma\circ \D_\Gamma$ when $\D_\Gamma$ is singular, the resulting expression is also valid for pure state models, as can be shown by a detailed analysis of the singular case ($\nu=1$). The QFI for such models is readily obtained 
\begin{align}\label{QFIscaledpure}
	I_Q=\frac12\frac{\nu^2}{1+\nu^2}\tr[\partial\Gamma\Gamma^{-1}\partial\Gamma\Gamma^{-1}]+2\partial d^\top\Gamma^{-1} \partial d.
\end{align}
First, notice that the contribution due to first moments is equal to that of Eq.~\eqref{classicalI}. This is due to the fact that there is always a reference frame for which the derivative $\partial d$ is along a set of mutually compatible variables. In addition, setting $(\nu=1)$ recovers the known expression for the QFI for pure state models~\cite{pinel_ultimate_2012}. Interestingly, the correction factor for thermal models $\nu^2/(1+\nu^2)$ approaches 1 in the large temperature limit ($\nu\gg1$), recovering the Fisher information contained in the Wigner distribution. As noncommutativity of the canonical variables dictates, a faithful sampling of the Wigner distribution is beyond reach except in the high temperature regime, when thermal fluctuations render quantum fluctuations irrelevant. This explains why, in the high temperature limit, $I_Q$ approaches the Fisher information of the Wigner distribution. In addition, notice that for pure state models with fixed first moments ($\partial d=0$), $I_Q$ is exactly 1/2 of the Fisher information contained in the state's Wigner function. This fact deserves further attention. Consider the the model $(\Gamma,\partial\Gamma)$ in the Williamson form
\begin{align}
	\Gamma_\theta= S_\theta \nu S_\theta^\top.
\end{align}
The essential quantity in Eq. \eqref{QFIscaledpure} is $W= S_\theta^{-1}\partial\Gamma_\theta S_\theta^{-1}{}^\top$, which is nothing but $\partial\Gamma_\theta$ expressed in the coordinates for which $\Gamma_\theta$ is diagonal. However, the canonical transformation $S_\theta$ is only determined up to an orthogonal symplectic transformation. Conveniently, $W$ is symmetric and Hamiltonian ($W\circ \omega=0$), thus there is always a symplectic orthogonal transformation $\mathcal O$ that diagonalizes it. Therefore, defining the symplectic transformation $T=\mathcal OS^{-1}$ and corresponding canonical coordinates $(\tilde Q_1,\ldots,\tilde Q_n,\tilde P_1,\ldots,\tilde P_n)=\tilde R=T R$, one has 
\begin{align}
	T \Gamma T^\top=\left(\begin{array}{cc}\nu & 0 \\0 & \nu\end{array}\right),\quad T\partial \Gamma T^\top=\left(\begin{array}{cc}\nu \lambda & 0 \\0 & -\nu \lambda\end{array}\right).
\end{align}
where $\lambda\geq0$ and we have made the $\nu$ dependency explicit in $T\partial\Gamma T^\top$. This illustrates a characteristic trait of all models of the kind \eqref{eq:Gibbs}, \emph{i.e.}, that there exist canonical coordinates $\tilde R^i$ for which variations of the parameter correspond to a collection of single-mode squeezing operations. In addition, it is clear that 
\begin{align}\label{eq:scaledQFI}
	I_Q=\frac{\nu^2}{1+\nu^2}\tr[\lambda^2].
\end{align}
Now consider the class of homodyne measurements obtained by measuring quadratures $\{\tilde Q_k\}$ (or $\{\tilde P_k\}$). The outcomes of such measurements are Gaussian distributed with covariance matrix $\gamma_\theta=\nu 1$ and $\partial\gamma_\theta=\pm \nu \lambda$, hence yielding $I_\cl$ given by
\begin{align}\label{Icl}
	I_\cl^\star=\frac{1}{2}\tr[\lambda^2].
\end{align}\nopagebreak
One immediately sees that for pure states ($\nu=1$) this is optimal, as follows from Eq.~\eqref{eq:scaledQFI}, $I_Q=I_\cl$. Hence, no other measurement can perform better. Let us pause for a moment to discuss what we mean by homodyne detection in this general multimode setting. In principle, we regard homodyne detection as the measurement of any set of compatible canonical variables, labelled $\{\hat Q_k\}$ for a suitably chosen canonical coordinates, $\hat R=TR$. However, the set of variables $\{\hat Q_k\}$ is only a convenient way to account for the information that the measurement outcomes provide. The real estimator $\Theta$ is some linear combination of the outcomes, and thus can be described as a linear combination of the canonical observables $\Theta=\hat\alpha^\top \hat R$, where $\alpha=(\alpha_1,\ldots,\alpha_n,0,\ldots,0)\in\mathbb{R}^{2n}$. Expressing $\Theta$ in terms of the original coordinates we get $\Theta=\alpha^\top R$, with $\alpha=T^\top \hat\alpha$. Hence, it remains to be seen that any such linear combination $\alpha^\top R$ can be implemented by simple linear combinations of the original $Q_k$ quadratures or passive Gaussian transformations thereof, and does not depend critically on the simultaneous measurement of incompatible variables or on some impractical active transformations. Parametrizing passive transformations as orthogonal symplectic matrices $V\in\M_{2n}$, we seek $V$ such that $\Theta=(V\alpha)^\top \,(V R)$, where $\tilde\alpha=V\alpha=(g,0)$. Writing $V$ with blocks $c, s\in\M_n$ we get
\begin{align}
	V\alpha=\left(\begin{array}{cc}c & s \\-s & c\end{array}\right)	\left(\begin{array}{c}
		\alpha_q \\
		\alpha_q
	\end{array}\right)
=
	\left(\begin{array}{c}
		c \alpha_q +s \alpha_p \\
		-s\alpha_q+c\alpha_p
	\end{array}\right),
\end{align}
so the condition that $[V\alpha]_p=-s\alpha_q+c\alpha_p=0$ is always achievable by \emph{e.g.}, $c=\mathrm{diag}({\alpha_{q;k}/|\alpha_{q;k}+i\alpha_{p;k}|})$ and $s=\mathrm{diag}({\alpha_{p;k}/|\alpha_{q;k}+i\alpha_{p;k}|})$. Let $g=c\alpha_q+s\alpha_p$, then $\Theta=\alpha^\top R=\sum_k  g_k \tilde Q_k$ [$\tilde R=VR$] shows that any linear combination of the canonical operators is implementable by passive transformations and homodyne detection on the original $n$ modes.

Going back to Eqs.~\eqref{eq:scaledQFI} and \eqref{Icl}, it is clear that for mixed states ($\nu>1$) there is, potentially, room for improvement, as $I_Q>I_\cl$. More general Gaussian measurements can be implemented by attaching a Gaussian ancilla and performing homodyne detection~\cite{giedke_characterization_2002, eisert_distilling_2002}. Adding an ancilla corresponds to replacing $\Gamma\rightarrow \Gamma\oplus \gamma$ and $\partial\Gamma\rightarrow\partial\Gamma\oplus0$, thus $\lambda\rightarrow \lambda\oplus 0$. Therefore, without loss of generality we can focus on homodyne detection. Consider the Fisher information $I_\cl$ of observables $\hat Q=\{\hat Q_1,\ldots,\hat Q_n\}$ out of a generic set of canonical coordinates $\hat R^i=\sum_jU^{ij}\tilde R^j$, parametrized by a symplectic matrix $U$
\begin{align}
	U=\left(\begin{array}{cc}a &b \\\cdot & \cdot\end{array}\right),\qquad ab^\top=ba^\top.
\end{align}
Outcomes $q=\{q_1,\ldots,q_n\}$ are distributed according to the covariance matrix $\hat\gamma=\nu [UU^\top]_{qq}=\nu(aa^\top+bb^\top)$, where the $[\,\cdot\,]_{qq}$ subindex indicates that the block corresponding to $\hat Q$ quadratures has to be taken. Likewise, the derivative $\partial\hat\gamma$ is of the form $\partial\hat\gamma=\nu(a\lambda a^\top-b\lambda b^\top)$. One can check that this model yields Fisher information
\begin{align}\label{homdyneFisher}
	I_\cl(a,b)=\frac{1}{2}\tr\Big[\big(\phi_a(\lambda)-\phi_b(\lambda)\big)^2\Big]
\end{align}
where $\phi_z(x)=\nu\, \hat\gamma^{-1/2} zxz^\top \hat\gamma^{-1/2}$, with $z\in\{a,b\}$. The maps $\phi_{a,b}$ are completely positive and $\phi=\phi_a+\phi_b$ is unital. Using the property $ab^\top=ba^\top$ one can show that $\phi$ is also trace-preserving. Finally, adding $2\tr[\phi_a(\lambda)\phi_b(\lambda)]\geq0$ to Eq.~\eqref{homdyneFisher} yields $I_\cl(a,b)\leq\frac{1}{2}\tr[\phi(\lambda)^2]$,
and using Kadison-Schwartz inequality one gets $I_\cl(a,b)\leq\frac{1}{2}\tr[\phi(\lambda^2)]=I_\cl^\star$ because $\phi$ is trace-preserving.

This shows that, for models of the form of Eq.~\eqref{eq:Gibbs}, with fixed first moments, homodyne detection of a suitable quadrature is always optimal among Gaussian measurements, with Fisher information upper bounded by $I_\cl^\star$, and heterodyne detection cannot improve its performance. Moreover, for pure models it is optimal in a wider sense --among all quantum measurements--. The optimality of homodyne detection generalizes and puts in context some earlier results~\cite{monras_optimal_2006,pinel_ultimate_2012}.

Further work can be envisaged in various directions. On the more practical side, identifying the most general class of models for which homodyne detection is optimal. On the theoretical side, application of these results to the study of Gaussian channels is a natural way of proceeding. In addition, one expects that some extension of the methods used here may be able to provide insight to paradigmatic non-Gaussian models such as phase diffusion and degaussified states.\\*

We thank H.~Cable, M.~Hayashi and B.-G.~Englert for useful discussions.  The Centre for Quantum Technologies is funded by the Singapore Ministry of Education and the National Research Foundation as part of the Research Centres of Excellence programme.

\bibliographystyle{apsrev4-1}	
\bibliography{gaussian_Fisher_arxiv}

\appendix

\section{The symmetric logarithmic derivative}\label{app:SLD}
We will make use of Einstein's summation convention. In addition, we will make a distinction between covariant and contravariant indices to bookkeep the transformation rules to which they comply to. In this spirit, the inverse of the symplectic metric $\omega$ is $\Omega_{ij}=[\omega^{-1}]_{ij}$, so that $\Omega_{ij}\omega^{jk}=\delta_i^k$. We give a different symbol to avoid confusion, because componentwise $\Omega_{ij}=-\omega^{ij}$. Also, we define $\partial_i\equiv\partial/\partial\xi^i$, and $\partial$ with no index refers to $\partial/\partial\theta$.

Taking the trace of Eq.~\eqref{eq:SLD} with the Weyl operators we get
\begin{align}\label{eq:sld2}
	\partial_\theta \chi_\rho(\xi)=\tr[\mathcal L_\theta\circ\rho\, W(\xi)].
\end{align}
As an ansatz, suppose $\mathcal L_\theta$ is at most quadratic in the canonical operators 
\begin{align}\label{eq:SLD2}
	\mathcal L_\theta=L^{(0)}+L^{(1)}_i R^i+L^{(2)}_{ij} R^i\circ R^j,
\end{align}
so the RHS of Eq.~\eqref{eq:sld2} can be written as
\begin{align}\nonumber
	\tr[\mathcal L_\theta\circ\rho\, W(\xi)]=&\,L^{(0)}+L^{(1)}_i \tr[\rho\,R^i\circ W(\xi)]\\
\label{eq:RHS}
		&+L^{(2)}_{ij}\tr[\rho\, (R^i\circ R^j)\circ W(\xi)].
\end{align}
On the other hand, the Gaussian characteristic function allows to write
\begin{align}\label{eq:derchi}
	\partial_\theta \chi_\rho(\xi)=\big(&i\xi \partial_\theta\bar d-\tfrac14\xi^\top \partial_\theta\bar \Gamma\xi\big)\chi_\rho(\xi).
\end{align}
where $\bar d$ and $\bar \Gamma$ in proper covariant-contravariant notation read
$\bar d_i=\Omega_{ij}d^j$ and $\bar \Gamma_{ij}=-\Omega_{ik}\Gamma^{kl}\Omega_{lj}$.

We begin by developping the RHS. The Weyl operators are defined as
\begin{align}
	W(\xi)=\exp(i\Omega(\xi,R)),
\end{align}
and we can show
\begin{align}
	\partial_iW(\xi)
		=&\,i\Omega_{ij}\, R^j\circ W(\xi),
\end{align}
or equivalently
\begin{align}\label{eq:RW}
	R^i\circ W(\xi)=-i\omega^{ij} \partial_j W(\xi)
\end{align}

It is tempting to define lower-indexed operartors $R_i=\Omega_{ij}R^j$ so that one can write $\partial_i W(\xi)= iR_i\circ W(\xi)$. However we will avoid this notation since the metric $\Omega$ is antisymmetric and it would be essential to be consistent on which index $R^j$ is contracted with. The second derivative reads
\begin{align}\label{eq:secder}
	\partial_i&\partial_j W(\xi)=-\Omega_{ik}\Omega_{jl}\,R^k\circ( R^l\circ W(\xi)).
\end{align}
The symmetric product is not associative $(A\circ B)\circ C\neq A\circ(B\circ C)$. To put Eq.~\eqref{eq:secder} in a manifestly symmetric form we use
\begin{align}
	R^k\circ(R^l\circ W(\xi))
	&=(R^k\circ R^l)\circ W(\xi)-\frac{1}{4}\xi^k\xi^lW(\xi)
\end{align}
so that Eq.~\eqref{eq:secder} reads
\begin{align}
	\partial_{ij} W(\xi)=-\Omega_{ik}\Omega_{jl}\left(R^k\circ R^l-\frac{1}{4}\xi^k\xi^l\right)\circ W(\xi)
\end{align}
which is manifestly symmetric under $i,j$. From this we can derive the relation
\begin{align}\label{eq:RRW}
	(R^k\circ R^l)\circ W(\xi)=&\left(\frac{1}{4}\xi^k\xi^l-\omega^{ki}\omega^{lj}\partial_{ij}\right)W(\xi).
\end{align}

Combining Eqs.~\eqref{eq:RW} and \eqref{eq:RRW} with Eq. \eqref{eq:RHS} we get
\begin{align}\label{eq:RHS2}
	\tr[&\rho\circ \mathcal L_\theta\, W(\xi)]\\
\nonumber
	&=\Big(L^{(0)}\!-\!iL^{(1)}_i\omega^{ik} \partial_k\!+\!L^{(2)}_{ij}\big(\tfrac{1}{4}\xi^i\xi^j-\omega^{ik}\omega^{jl}\partial_{kl}\big)\Big)\chi_\rho(\xi).
\end{align}
Using Eqs.~\eqref{eq:derchi} and \eqref{eq:RHS2}, Eqs.~\eqref{eq:sld2} reads
\begin{align}\label{eq:SLD3}
	\big(&i\xi \partial\bar d-\tfrac14\xi^\top \partial\bar \Gamma\xi\big)\chi_\rho(\xi)\\
\nonumber	&=\Big(L^{(0)}-iL^{(1)}_i\omega^{ik} \partial_k +L^{(2)}_{ij}\big(\tfrac{1}{4}\xi^i\xi^j-\omega^{ik}\omega^{jl}\partial_{kl}\big)\Big)\chi_\rho(\xi).
\end{align}
It is straightforward to evaluate the derivatives of $\chi_\rho$. These are
\begin{align}
	-iL^{(1)}_i\omega^{ik}\partial_k\chi_\rho(\xi)
	&=\big(L^{(1)}_id^i-\tfrac{i}2 L^{(1)}_i\Gamma^{ik}\Omega_{kl}\xi^l\big)\chi_\rho(\xi)
\end{align}
and
\begin{align}
	L^{(2)}_{ij}\big(&\tfrac{1}{4}\xi^i\xi^j-\omega^{ik}\omega^{jl}\partial_{kl}\big)\chi_\rho(\xi)=\\
\nonumber
	=&\Big(L^{(2)}_{ij}(d^id^j+\tfrac12 \Gamma^{ij}-\tfrac{i}2L^{(2)}_{ij}(d^i\Gamma^{jk}+d^j\Gamma^{ik})\Omega_{kl}\xi^l\\
\nonumber
	&~~-\tfrac{1}{4}L^{(2)}_{ij}(\Gamma^{ik}\Gamma^{jl}-\omega^{ik}\omega^{jl})\Omega_{kr}\Omega_{ls}\xi^r\xi^s\Big)\chi_\rho(\xi).
\end{align}
Rewriting Eq.~\eqref{eq:SLD3}, and recalling that $\chi_\rho$ is nowhere zero, we get 
\begin{align}\label{eq:SLD4}
	i\xi \partial\bar d\,-\,&\tfrac14\xi^\top \partial\bar \Gamma\xi=\\
\nonumber	
=&\,L^{(0)}+L^{(1)}_id^i+L^{(2)}_{ij}(d^id^j+\tfrac12 \Gamma^{ij})\\
\nonumber	
			 &-\tfrac{i}2\big(L^{(1)}_i\Gamma^{ik}+L^{(2)}_{ij}(d^i\Gamma^{jk}+d^j\Gamma^{ik})\big)\Omega_{kl}\xi^l\\
\nonumber	
			 &-\tfrac{1}{4}L^{(2)}_{ij}(\Gamma^{ik}\Gamma^{jl}-\omega^{ik}\omega^{jl})\Omega_{kr}\Omega_{ls}\xi^r\xi^s,
\end{align}
Equaling the different orders in $\xi$ we get
\begin{subequations}\label{eq:L012}
\begin{align}
	L^{(0)}+L^{(1)}d+d^\top L^{(2)} d+\tfrac12 \tr[L^{(2)}\Gamma]&=0\\
	L^{(1)}+(L^{(2)}+L^{(2)}{}^\top) d&=2\Gamma^{-1}\partial d\\
\label{eq:L2}	
	\Gamma L^{(2)}\Gamma-\omega L^{(2)}\omega^\top&=\partial\Gamma.
\end{align}
\end{subequations}
By defining the linear map $\D_X$ as 
\begin{align}
	\D_X(Y)=X YX^\top-\omega Y\omega^\top, 
\end{align}
Eq.~\eqref{eq:L2} can be written as
\begin{align}
	 \D_\Gamma(L^{(2)})=\partial\Gamma,
\end{align}
and a solution exists when $\partial\Gamma$ lies within the range of $\D_\Gamma$. Using $\D_\Gamma^{-1}$ to denote the pseudoinverse of $\D_\Gamma$ and the fact that $\D_Y(X^\top)=\D_Y(X)^\top$ we rewrite Eqs.~\eqref{eq:L012} as
\begin{subequations}
\begin{align}
	L^{(2)}=\,&\D_\Gamma^{-1}(\partial \Gamma)\\
	L^{(1)}=\,&2\Gamma^{-1}\partial d-2\D_\Gamma^{-1}(\partial\Gamma)d\\
	L^{(0)}=\,&d^\top \D_\Gamma^{-1}(\partial\Gamma)d-2d^\top\Gamma^{-1}\partial d-\tfrac12\tr[\D_\Gamma^{-1}(\partial\Gamma)\Gamma].
\end{align}
\end{subequations}
Putting this back into Eq.~\eqref{eq:SLD2} we get
\begin{align}
	\mathcal L_\theta=&\,\D^{-1}_\Gamma(\partial\Gamma)_{ij} (R^i-d^i)\circ (R^j-d^j)\\
	&+2\partial d^i \Gamma^{-1}_{ij}(R^j-d^j)\\
	&-\frac{1}{2}\tr[\D^{-1}_\Gamma(\partial\Gamma)\Gamma]
\end{align}
This proves Eqs.~\eqref{eq:SLDfinal} and \eqref{SLDparams}.

As suggested above $\D_\Gamma$ may be singular. To analyze the spectrum of $\D_\Gamma$ it is convenient to consider $\Gamma=S\Gamma_\th S^\top$ in Williamson form, where $\Gamma_\th$ is a (thermal) Gibbs state. In this representation, 
\begin{align}\label{eq:Dthermal}
	\D_\Gamma=(S\otimes S)\D_{\Gamma_\th}(S\otimes S)^\top
\end{align}
and the eigenvalue equation $\D_{\Gamma_\th}(X)=\lambda X$ reads $\Gamma_\th X \Gamma_\th-\omega X \omega^\top=\lambda X$. Since $\Gamma_\th$ is diagonal and proportional to the identity in each $2\times2$ block corresponding to each mode, the eigenvalue equation decouples into $n^2$ independent equations, labelled by the pair $(i,j)$, corresponding to the $i$-th and $j$-th modes in the row and column resp. 
\begin{align}\label{eq:decoupledeigenvalues}
	\omega X_{ij} \omega^\top = (\nu_i \nu_j-\lambda) X_{ij}.
\end{align}
The map $\omega \cdot \omega^\top$ is an involution, thus its eigenvalues are $\pm1$. Hence, the eigenvalues of $\D_{\Gamma_\th}$ are of the form $\lambda=\nu_i\nu_j\mp1$, where $\nu_i$ are the symplectic eigenvalues of $\Gamma$. In addition, matrices $1$ and $\omega$ are eigenvectors of ${\omega\,\cdot\,\omega^\top}$ corresponding to eigenvalue $+1$, and Pauli matrices $\sigma_x$ and $\sigma_z$ correspond to eigenvalue~$-1$. Hence $1,\omega$ are eigenvectors of $\D_{\Gamma_\th}$ with eigenvalue $\nu_i\nu_j-1$, and $\sigma_x,\sigma_z$ are associated to $\nu_i\nu_j+1$.
Thus, the spectrum of $\D_{\Gamma_\th}$ is given by $\lambda_{ij}^\pm=\nu_i\nu_j\mp1$. In particular, one can see that if $\Gamma$ has symplectic eigenvalues equal to 1 (or $\Gamma+i\omega$ is singular), then $\D_{\Gamma_\th}$ is singular, and so is $\D_\Gamma$. The kernel of $\D_{\Gamma_\th}$ consists of sparse matrices populated only in blocks $(i,j)$ such that $\nu_i=\nu_j=1$, with entries of the form $\alpha_{ij}1+\beta_{ij}\omega$,
\begin{align}\label{eq:kernel}
	\ker(\D_{\Gamma_\th})=\left(\begin{array}{cccccc}
		0		&\cdots	&0							&\cdots	&0\\
		\vdots	&\ddots	&\vdots						&		&\vdots\\
		\vdots	&\cdots	&\alpha1+\beta\omega	&\cdots	&0\\
		\vdots	&		&\vdots						&\ddots	&\vdots\\
		0		&\cdots	&0							&\cdots	&0
	\end{array}\right).
\end{align}
Thus $\D_\Gamma$ is nonsingular only when the symplectic eigenvalues of $\Gamma$ are strictly greater than 1, and if $\D_\Gamma$ is singular, its kernel consists of matrices with with nonzero entries in blocks corresponding to vacuum modes in the Williamson decomposition of $\rho$. 

\section{The quantum Fisher information}\label{app:QFI}
The QFI is defined as $I_Q=\tr[\rho_\theta \mathcal L_\theta^2]$. Using the notation introduced in the text, a map $X\rightarrow A X B^\top$ reads $A\otimes B$ and the action is specified as follows
\begin{align}
	A\otimes B|X)=|A X B^\top).
\end{align}
With this $\D_X=X\otimes X-\omega\otimes\omega$ has 4 upper indices $[\D_X]^{ijkl}=X^{ik}X^{jl}-\omega^{ik}\omega^{jl}$, of which the last two are contracted with the argument matrix,
\begin{align}
\nonumber
	[\D_X(Y)]^{ij}&=[\D_X]^{ijkl}Y_{kl}\\
\nonumber
		&=\big(X^{ik}X^{jl}-\omega^{ik}\omega^{jl}\big)Y_{kl}\\
\nonumber
		&=X^{ik}Y_{kl}X^\top{}^{lj}-\omega^{ik}Y_{kl}\omega^\top{}^{lj}\\
		&=[XYX^\top-\omega Y\omega^\top]^{ij}.
\end{align}
On the other hand, supposing that $\D_X$ is nonsingular, $[\D_X^{-1}]_{ijkl}[\D_X]^{klrs}Y_{rs}=Y_{ij}$, and
\begin{align}
	[\D_X^{-1}(Y)]_{ij}=[\D_X^{-1}]_{ijkl}Y^{kl},
\end{align}
and 
\begin{align}
	[\D_X^{-1}]_{ijkl}[\D_X]^{klrs}=\delta^r_i\delta^s_j.
\end{align}
\begin{widetext}
We will use the relation~\cite{monras_information_2010}
\begin{align} 
	\tr[\rho_\theta (\hat R^i\,\circ \,\hat R^j)\circ(\hat R^k\circ \hat R^l)]=\frac{1}{4}\left[\Gamma^{ij}\Gamma^{kl}+[\D_\Gamma]^{ijk'l'}(\delta^k_{k'}\delta^l_{l'}+\delta^k_{l'}\delta^l_{k'})\right],
\end{align}
where the centered canonical operators $\hat R=R-d$ have been defined. With this, $I_Q$ reads
\begin{subequations}
\begin{align}
	I_Q=&\,
		\D_\Gamma^{-1}(\partial\Gamma)_{ij}\D_\Gamma^{-1}(\partial\Gamma)_{kl}\tr[\rho_\theta (\hat R^i\circ \hat R^j)\circ(\hat R^k\circ \hat R^l)]\\
\nonumber
		&+\left(4[\Gamma^{-1}\partial d]_i[\Gamma^{-1}\partial d]_j-\tr[\D^{-1}_\Gamma(\partial\Gamma)\Gamma]\D_\Gamma^{-1}(\partial\Gamma)_{ij}\right)
		\tr[\rho_\theta\,\hat R^i\circ \hat R^j]+\frac{1}{4}\tr[\D^{-1}_\Gamma(\partial\Gamma)\Gamma]^2\\
	=&\,\frac{1}{4}\D_\Gamma^{-1}(\partial\Gamma)_{ij}\D_\Gamma^{-1}(\partial\Gamma)_{kl}[\D_\Gamma]^{ijk'l'}(\delta^k_{k'}\delta^l_{l'}+\delta^k_{l'}\delta^l_{k'})+4[\Gamma^{-1}\partial d]_i[\Gamma^{-1}\partial d]_j\tr[\rho_\theta\,\hat R^i\circ \hat R^j]\\
	=&\,\frac{1}{4}\D_\Gamma^{-1}(\partial\Gamma)_{ij}\D_\Gamma^{-1}(\partial\Gamma)_{kl}[\D_\Gamma]^{ijk'l'}(\delta^k_{k'}\delta^l_{l'}+\delta^k_{l'}\delta^l_{k'})+2\partial d^\top \Gamma^{-1}\partial d.
\end{align}
\end{subequations}
\end{widetext}
Since $\D_\Gamma^{-1}(\partial\Gamma)$ is symmetric, we can write
\begin{align}
\nonumber
	I_Q&=\frac{1}{2}\D_\Gamma^{-1}(\partial\Gamma)_{ij}[\D_\Gamma]^{ijkl}\D_\Gamma^{-1}(\partial\Gamma)_{kl}\\
	&+2\partial d^\top \Gamma^{-1}\partial d.
\end{align}
Finally, notice that 
\begin{align}
	[\D_\Gamma]^{ijkl}[\D_\Gamma^{-1}(\partial\Gamma)]_{kl}={[\D_\Gamma\circ \D_\Gamma^{-1}(\partial\Gamma)]^{ij}}=[\P(\partial\Gamma)]^{ij}
\end{align}
where $\P=\D_\Gamma\circ \D_\Gamma^{-1}$ is the projector onto the support of $\D_\Gamma$. Hence,
\begin{align}
	I_Q=\frac{1}{2}\tr[\D^{-1}_\Gamma(\partial\Gamma) \,\P(\partial\Gamma)]+2\partial d^\top \Gamma^{-1}\partial d
\end{align}
To conclude, notice that $\D_\Gamma$ is self adjoint, \emph{i.e.}, $\tr[\D_\Gamma(X)^\top Y]=\tr[X^\top \D_\Gamma(Y)]$, as a consequence of $\Gamma$ being symmetric. So is $\D_\Gamma^{-1}$. Also, the presence of $\D_\Gamma^{-1}$ makes $\P$ unnecessary, i.e., ${\D_\Gamma^{-1}\circ \P=\D_\Gamma^{-1}}$, hence we can write
\begin{align}
	I_Q=\frac12\tr[\partial\Gamma\,\D_\Gamma^{-1}(\partial\Gamma)]+2\partial d^\top\Gamma^{-1}\partial d
\end{align}
or, in  braket notation,
\begin{align}
	I_Q=\frac12 (\partial\Gamma|\D_\Gamma^{-1}|\partial\Gamma)+2\partial d^\top\Gamma^{-1}\partial d.
\end{align}
More explicitly,
\begin{align}
	I_Q=\frac12 (\partial\Gamma|(\Gamma\!\otimes\!\Gamma-\omega\!\otimes\!\omega)^{-1}|\partial\Gamma)+2\partial d^\top\Gamma^{-1}\partial d,
\end{align}
which is the expression reported in Eq.~\eqref{eq:QFI}.

\end{document}